\documentclass[12pt]{iopart}

\usepackage{amssymb}

\usepackage{txfonts}

\usepackage[pdftex,colorlinks=true,bookmarks=false,citecolor=blue,urlcolor=blue]{hyperref} 

\usepackage{graphicx}
\usepackage{dcolumn}
\usepackage{bm}

\usepackage{multicol} 

\usepackage{CJK}

\begin{document}


\title[Transition from ultrafast laser photo-electron emission to space-charge-limited current ....]
{Transition from ultrafast laser photo-electron emission to space-charge-limited current in a 1D gap}

\author{Liu Yangjie()$^{1}$\footnote{Visiting student in Department of Physics, Soochow University, Suzhou 215006 China. Contact \mailto{liuy0074@e.ntu.edu.sg}.}, L K Ang$^{2,*}$}
\address{$^1$ School of Electrical and Electronic Engineering, \\
Nanyang Technological University, Singapore 639798.}


\address{$^2$ Engineering Product Development, \\
Singapore University of Technology and Design, Singapore 138682.}
\ead{ $^{*}$ \mailto{Correspondence author--ricky\_ang@sutd.edu.sg}}

\begin{abstract}
\noindent 
A one-dimensional (1D) model has been constructed to study the transition of the time-dependent ultrafast laser photo-electron emission from a flat metallic surface to the space-charge-limited (SCL) current, including the effect of non-equilibrium laser heating on metals at the ultrafast time scale. 
At a high laser field, it is found that the space charge effect cannot be ignored and the SCL current emission is reached at a lower value predicted by a short-pulse SCL current model that assumed a time-independent emission process.
The threshold of the laser field to reach the SCL regime is determined over a wide range of operating parameters.
The calculated results agree well with particle-in-cell (PIC) simulation.
It is found that the space charge effect is more important for materials with lower work function like tungsten (4.4 eV) as compared to gold (5.4 eV). However for a flat surface, both materials will reach the SCL regime at the sufficiently high laser field such as $> 5 \rm {GV \, m^{-1}}$ with a laser pulse length of tens to a hundred femtoseconds.
\end{abstract}










\footnote 
{Rejected by {\itshape Phys. Plasmas}(05-Oct-2013),\\
 {\JPD} \\Received (11-Oct), moderate revision suggested(08-Jan-2014), re-submitted 16-Jan (Article reference: JPhysD-100502.R1), 
 Accepted for publication 27-Jan, proofread 15-Feb, 
 Published online 6-March-2014. \\
 Please cite this Paper as: J. Phys. D: Appl. Phys. 47 (2014) 125502. \url{http://iopscience.iop.org/0022-3727/47/12/125502/}}

\maketitle


\section{Introduction}

For electrons emitted from a surface into a free space, the amount of current emitted at low-current regime is dominated by the emission mechanism, which is known as source-limited emission.
The mechanism can be divided into three types, namely thermionic emission, field emission, and photoemission, which is respectively, described by 
the Richardson-Laue-Dushman (RLD) law \cite{RichEEHB}, the Fowler-Nordheim (FN) law \cite{Fow1928}, and the Fowler-Dubridge (FD) law \cite{Fow31,Dub32,Dub33}.
A good overview can be found in a recent paper by Jensen \cite{Jen07AIEP}. 
In particular, all three emission mechanisms can be combined in a generalized model \cite{Jen02}.

At high-current regime, the amount of the emitted current will be influenced by the space charge effect, and it is known as the space-charge-limited (SCL) current, which
describes the maximum current density allowed for steady-state electrons emitted from the cathode and transported across the gap. 
For a one-dimensional (1D) gap of spacing $D$ and a dc voltage of $V_g$ , the SCL current is governed by the1D classical Child-Langmuir (CL) law \cite{Child1911,Langmuir1913}, given by
\begin{equation}\label{JCL}
J_{\rm CL}=\frac{4\epsilon_0}{9D^2}\sqrt{\frac{2e}{m}}V_g^{3/2},
\end{equation}
where $\epsilon_0$, $e$ and $m$ is, respectively, vacuum permittivity, electron charge and electron mass. 
SCL electron flow occurs when the charge of the emitted electrons is sufficient to suppress the electric field at the cathode to zero, and the electrostatic potential distribution function $\phi(x)$ is  
\begin{equation}\label{Vxdis}
\phi(x)=V_g\left(\frac{x}{D}\right)^{4/3}.
\end{equation}

The transition from source-limited emission to SCL current is important in the development of cathodes such as field emitters and photocathode, which is, respectively, based on field emission and photoemission.
The transition from the field emission to SCL current has been developed for large \cite{Lau94} and small \cite{Koh06} gaps.
For the transition from photoemission to SCL current, the CL law can not be used directly, because the pulse length of the electron photoemission is normally much smaller than the electron transit time across the gap.
This short-pulse effect on the CL law had been developed in a 1D classical model \cite{Valfells2002} given by
\begin{equation} \label{spJCL}
J_{\rm crit}=2\frac{1-\sqrt{1-\frac{3}{4}X_{\rm CL}^2}}{X_{\rm CL}^3} J_{\rm CL},
\end{equation}
where $X_{\rm CL}= \tau_{\rm p} / \tau (< 1)$ is the normalized pulse length and $\tau = 3D\sqrt{m/{2eV_g}}$ is the transit time at the SCL condition.
A more recent model has also been developed to include the quantum effects when the electron de Broglie wavelength is comparable or smaller than the electron pulse width \cite{Ang2007}.

Recently, significant efforts have been made in using ultrafast laser to induce electron photo-field emission from metallic tips~\cite{Hommelhoff2006a,Hommelhoff2006b,Ropers2007,Barwick2007,Wu2008,Sch2010,Yanagisawa2011,Yalunin2011,Pant2012,Ang2013,Bormann2010,Kruger2011,Herink2012,Park2012,Mueller2013,Pant2013}.
In most studies, space charge effect has been ignored, which may be important at the high-current regime operating at the high laser fields \cite{Wendelen2010, Wendelen2012}.
Depending on the operating conditions, the emission process due to ultrafast laser excitation can be complicated, which leads to multiphoton emission \cite{Ropers2007}, optical tunneling \cite{Hommelhoff2006a}, strong field photoemission \cite{Yalunin2011}, and above threshold photoemission \cite{Sch2010}. The onset of the optical tunneling from the multiphoton emission can be determined by a formula \cite{Pant2012}, which is at about 9.81 $\rm V\, nm^{-1}$ for an 8 femtosecond laser pulse.

In this paper, we only focus on the photoemission process, and are interested to develop a simple 1D model to show the transition from the ultrafast-laser-induced multi-photon emission to the space charge limit at the ultrafast time scale.
Due to the long relaxation time scale (ps), the multiphoton emission process is governed by the non-equilibirum heating model \cite{Wu2008}, which had shown that the emission is time-dependent at a time scale less than 1 ps for metals.
Note this phenomenon was verified by an experiment in 2011 \cite{Yanagisawa2011}.

While \eref{spJCL} is able to account for the effect of short pulse, yet it has assumed that the current injected into the gap is time-independent.
Here, from $t = 0$ (beginning of the laser pulse) to $t = \tau_{\rm p}$ (end of the pulse), the injected current density from the cathode at $x = 0$ is emitted as a function of time, given by $J(t)$.
To solve this problem, we need to find the spatial variation of the injected electron density into the gap, $J(x)$ for $0 \leq x \leq s$ , where $s$ ( $<< D$) is the position of the beam front at at $t = \tau_{\rm p}$.
Once this $J(x)$ is obtained, we may use the similar approach in the short pulse model \cite{Valfells2002} to calculate the space charge limited electron flow for the ultrafast-laser-induced photo-field emission.

It is important to note that the model presented here has ignored the sharpness of the tip. Thus the prediction can be considered only as a zero-order estimation.  
While  the 2D CL law has been developed~\cite{Lau2001}, it can not be applied directly as the emitting area (in a sharp tip) is much smaller than the gap spacing.
The model is a quantitative one that is able to compare experiments in using ultrafast laser to excite multi-photon electron emission from a flat surface.
In future, it can be extended to a non-uniform and at least a 2D model to account for the sharpness of the tip in order to compare quantitatively with the experiments using sharp tips.
Note the latter is not a trivial task, as there is no protrusive CL law which is valid for even steady-state electron emission from a sharp tip.


\section{Model}
Consider a gap  of spacing $D$ with an external fixed electric field $F_{\rm dc} = V_g / D$.
The cathode (at $x$ = 0) is excited by an ultrafast laser to induce electron emission with a time-varying current density $J(t)$.
The emitted electrons will travel under the influence of the electric field  (including space charge field) to reach a distance of $s$ 
($< D$) within the laser pulse-length $\tau_{\rm p}$, which is smaller than the transit time $\tau$. 
Note the laser profile is a step-like function with a temporal duration of $\tau_p$.

In the region of $0 \leq x \leq s$, the electrostatic potential is
\begin{equation}
\label{phiwhole}
\phi (0\leq x\leq s)
=F_{\rm dc}x+\Delta\phi(x),
\end{equation}
where $\Delta\phi(x)$ is the space-charge electrostatic potential.
The value of $s$ can be calculated by solving the equation of motion for $x(t)$ given by
\begin{eqnarray}\label{xt}
\frac{\rm d^2}{{\rm d}t^2}x(t)
&=&\frac em[F_{\rm dc}+\frac{\rm d}{{\rm d}x}\Delta\phi(x)],
\end{eqnarray}
with initial conditions: $x(t=0)=0$ and $x'(t=0)=v_0 \approx 0$, where the primed symbol indicates the first derivative. 
Here, the initial velocity $v_0$ is kept to be reasonably small (but not equal to zero) to avoid the difficulty in the numerical integration of Poisson equation (see below) at $x$ = 0.
In general, we have $v_0\tau_{\rm p}\ll s$ in our calculation.

Near the cathode surface, we consider a surface potential barrier,  
\begin{equation}\label{newV}
V(x)=V_{\rm f}+\Phi-\frac{e^2}{16\pi \epsilon_0x}-exF_{\rm dc}-e\Delta\phi(x).
\end{equation}
where $V_{\rm f}$ is the Fermi energy, $\Phi$ its work function and the third term the classical image charge potential.
Using \eref{newV}, we obtain that the time-dependent tunneling current density $J(t)$, which is calculated from using formula
\begin{equation}\label{J}
J(t)=\frac{em}{2\pi^2\hbar^3}\int_0^{\infty}{\rm d}W T(W)\int_W^{\infty}f(E,t){\rm d}E.
\end{equation}
Here, $T(W)$ is the electron tunneling probability at energy level $W$ through the surface potential barrier (cf. \eref{newV}), which is based on the modified Wentzel-Kramers-Brillouin (WKB) method, 
$f(E,t)$ is the time-varying non-equilibrium electron distribution function based on our previous model \cite{Wu2008}, $W$ is a dummy variable in the integration, and $\hbar$  is the reduced Plank constant.
The details on how to obtain $T(W)$ and $f(E,t)$ can be found in Ref. [18] and [22].

Using \eref{J}, the emitted charge density (per unit area) within the laser pulse-width is calculated by
\begin{equation}\label{NEC}
\sigma_{\rm EC}=\int_0^{\tau_{\rm p}}{J(t)}{\rm d}t.
\end{equation}
To determine $\Delta\phi(x)$ in the region $0 \leq x \leq s$, we solve the Poisson equation , 
\begin{equation} \label{P}
\frac{d^2}{dx^2} \Delta\phi(x)= \frac{J(x)}{\epsilon_0 v(x)},
\end{equation}
where $v(x)$ is the velocity profile of the electron flow that can be obtained using the energy conservation,
\begin{equation}
\frac{mv^2(x)}{2} = e[F_{dc}x+\Delta\phi(x)].
\end{equation}
Here, $J(x)$ is the current density profile obtained at the end of the pulse at $t = \tau_p$, which is related to \eref{J}, and it can be expressed as 
\begin{equation}\label{J2}
{J(x)}=J[\tau_{\rm p}-t(x)],
\end{equation}
where $t(x)$ is the inverse function of $x(t)$ solved in \eref{xt}.
Combining \eref{P} to \eref{J2}, we have
\begin{eqnarray}\label{Poisson}
\frac{{\rm d}^2}{{\rm d} x^2}\Delta\phi(x)
&=&\frac{J[\tau_{\rm p}-t(x)]}{\sqrt{\frac{2e}{m}(F_{\rm dc}x+\Delta\phi(x))}}.
\end{eqnarray}

The boundary conditions for solving \eref{Poisson} are the zero space potential at the cathode ($x$ = 0) and the continuous electric field at the beam front ($x = s$), which are respectively, 
\begin{equation} 
\Delta\phi(0)=0, 
\end{equation}
and 
\begin{eqnarray}\label{sboundary} 
\frac{\rm d}{{\rm d} x}\Delta\phi(x)\vert_{x=s}=-\frac{\Delta\phi(s)}{D-s}. 
\end{eqnarray}
Note \eref{sboundary} is reduced from ${\rm d\phi(x)}/{{\rm d} x}\vert_{x=s} = (V_g-\phi(s))/(D-s)$.
Finally, the electrostatic potential in the vacuum region of $s < x \leq D$ (in front of the electron beam) is
\begin{equation}
\Delta\phi(s\leq x\leq D)=\frac{\Delta\phi(x=s)}{D-s} (D-x).
\end{equation}

From the equations above, it is required to solve $\Delta\phi(x)$, $s$ and $x(t)$ numerically to determine the SCL current density $J(t)$ or the SCL charge density $\sigma_{EC}$ consistently. 
Here we construct  a {\emph {numerical}} algorithm to perform the calculation iteratively.

On the first step, we assume $\Delta\phi_1(x)=0$ in \eref{newV} to obtain the electron current density (without space charge effects) $J_1(t)$ from \eref{J}, which gives the first estimated value for $J(t)$.
By substituting $\Delta\phi=0$ into \eref{xt}, we also obtain the first estimated time-profile $x(t)$ and hence its inverse function $t(x)$. Here the value of $s$ is determined by $s = x$ ($t=\tau_p$).
With this $J_1(t)$, we can solve the Poisson equation from \eref{Poisson} for the new $\Delta\phi_2(x)$, which can be used to estimate the new current density $J_2(t)$.
Here, $J_2(t)$ is different from $J_1(t)$ due to the finite value of $\Delta\phi_2(x)$ as the space charge effect has been included.
The iterative process will continue until a convergence is reached which is determined by  ${\vert J_{i+1}(t=\tau_{\rm p})-J_i(t=\tau_{\rm p})\vert}/{J_i(t=\tau_{\rm p})}\leq 1.5\times 10^{-4}$, where $i$ is the iterative step number.
Once the convergence is reached, we can determine the space charge potential $\Delta\phi(x)$, and compute the SCL charge density $\sigma_{EC}$ according to \eref{NEC}, which can be compared to our previous work that excluded the space charge effects completely \cite{Wu2008}.

It is important to note that this approach has ignored completely the space charge field within the short time scale less than $\tau_p$. The approach is similar to the short pulse model with constant  current density \cite{Valfells2002}, which had been confirmed with particle-in-cell (PIC) code. 
Our results will also be compared with PIC simulation with a time-dependent emission current, which shows rather good agreement.

Before presenting the results, we are interested to calculate the saturation of the SCL current density at high fields due to the suppression of the $total$ electric field towards zero at the cathode.
Thus the space charge potential $\Delta\phi(x)$ must be large enough to suppress the applied DC field $F_{\rm dc}$, given by 
\begin{equation}\label{limit}
F_{\rm dc}=-\frac{\rm d}{{\rm d}x}\Delta\phi\vert x=0.
\end{equation}
When the condition above is fulfilled, we define a critical  (SCL) current density ,
 \begin{equation} 
 \label{JSCL}
J_{\rm SCL}(t)=fJ_0(t),   
\end{equation}
where $f$ is an enhancement factor over the current density $J_0$ obtained from the time-dependent ultrafast laser emission model (without space charge effects) ~\cite{Wu2008}.
Here, $f$ can be expressed by
\begin{equation}\label{f}
f=-\frac{F_{\rm dc}}{\frac{\rm d}{{\rm d}x}\Delta\phi_0(x=0)},
\end{equation}
where the $\Delta\phi_0$ is the space charge potential obtained by solving the Poisson equation using $J_0(t)$.

\section{Results and Discussions}

Here, we will not present the calculated $f(E,t)$ in this paper, which is due to a step-function like ultrafast profile, and it has been shown elsewhere (see figure 1 in Ref. [18]). The property of the $f(E,t)$ for a Gaussian-like ultrafast laser profile can also be found in a recent paper (see figure 1 in Ref. [22]).  From these calculated $f(E,t)$, we solve (6), (7) and (12) to obtain a self-consistent emission density including the space charge effect, which is linked to the emission process and also the non-equilibrium laser-metal interaction at the ultrafast time scale. A more recent model on the non-equilibrium laser-metal interaction can also be found in Ref. [28].

\begin{figure*}[hc]
 \includegraphics[width=1.1\textwidth]{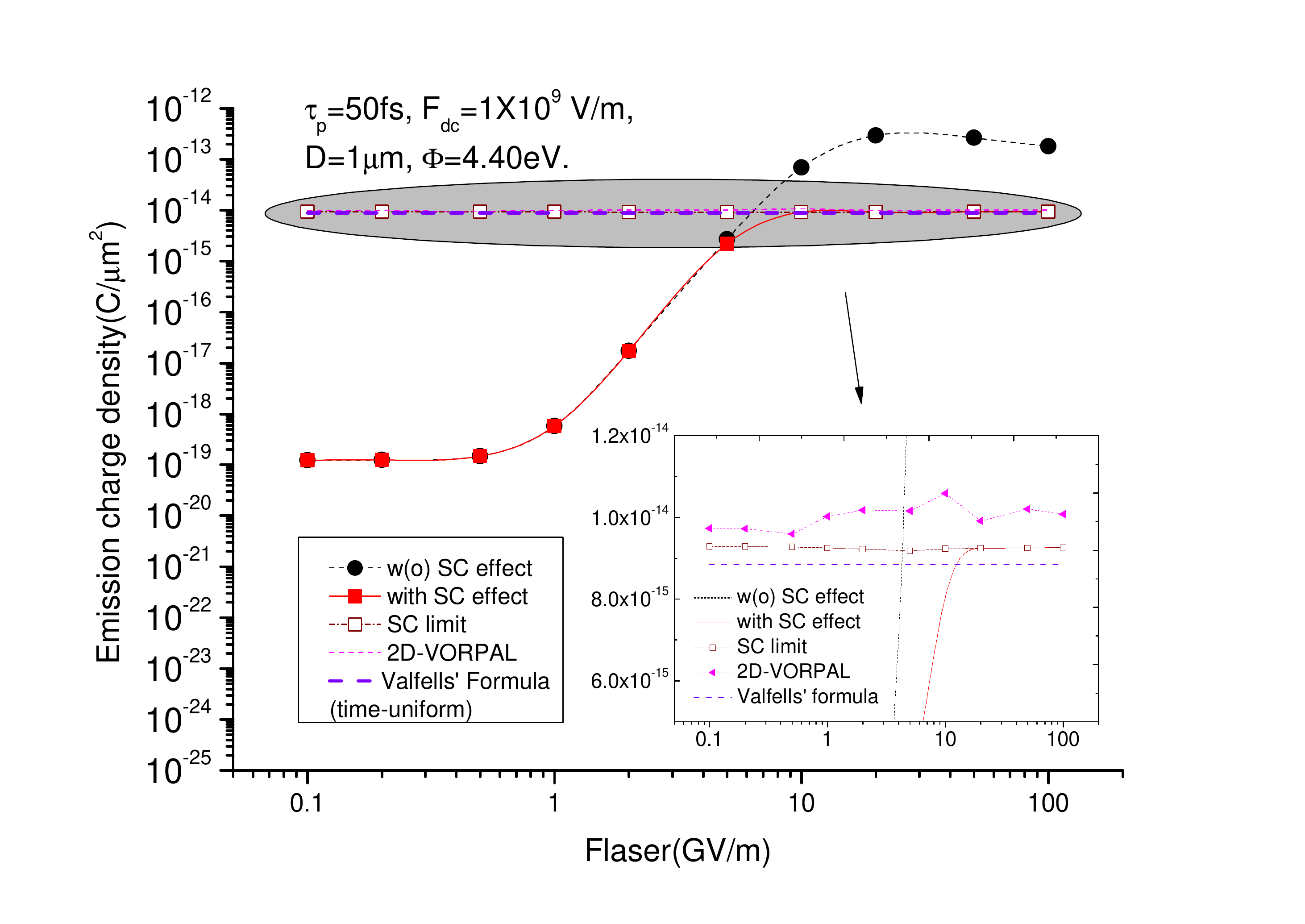}

\caption { \label{fig:Fig1} (Color online)
Emission charge density $\sigma_{EC}$ as a function of laser field $F_L$ for cases with (red solid square) and without the space charge effect (black dot) at $\tau_{\rm p}=50{\rm fs}$, $F_{\rm dc}=1\rm GV \;m^{-1}$, $D = 1 {\rm \mu m}$, $\Phi$ = 4.4 eV.
 The comparisons are the space charge (SC) limit from ~\eref{JSCL}, Valfells' formula \cite{Valfells2002} and 2D VORPAL (PIC simulation results).
(Inset) Enlargement of the shaded region near the SCL regime.
}

\end{figure*}

In figure 1, we present the emitting charge density $\sigma_{EC}$ as a function of the laser field $F_{\rm L}$ with the following parameters:
$\tau_{\rm p}$ = 50 fs, $F_{\rm dc} = 1 {\rm GV\; m^{-1}}$, $D= 1  {\rm \mu m}$  and $\Phi = 4.4 \rm eV$. In this figure, we see that at the low laser field $F_L < 5 {\rm GV\;m^{-1}}$, both $\sigma_{\rm EC}$ values calculated by the models with (red squares) and without space-charge effect (black circles) are nearly identical, which indicates space-charge effects are not important in this range of the low laser field $F_{\rm L}$. 
Around $F_{\rm L} = 5 {\rm GV\;m^{-1}}$, we observe a smooth transition into the SCL regime (space-charge limits indicated in open square) (also cf. the inset of figure 1 which zooms in the shaded oven zone). 
In comparison with the time-independent model (dashed purple line) by Valfells~\cite{Valfells2002}, our results (time-dependent) of SC limit are slightly higher. 

Above the threshold laser field, increasing laser power will not further increase the emitted charge, and the limit is determined by the SCL current.
At this regime, the amount of charge density remains as a constant, which is simply controlled by $Q = C \cdot V_g$, where $C$ is is the capacitance per area of the gap.

To compare our results with PIC simulation, we use a 2D PIC code called VORPAL~\cite{Nieter2004}.
In the simulation, we inject a time-dependent electron current density based on our model \cite{Wu2008}, and we determine the SCL current when reflection of electrons is detected \cite{Mahalingam2009, Chen2011}. 
We have a large emitting area so that the electron flow has a uniform space distribution in the transverse directions
in order to be comparable with our 1D model.
The comparison shows that our model (open square) is slightly lower than the PIC simulation results (triangle) as shown in the inset of figure 1. Because both our method and PIC simulation give higher emission densities than Valfells' formula does, this demonstrates or at least inspires one to ponder whether it is possible to achieve a higher upper limit of space-charge limited emission density in time-dependent injection current case\footnote{Also cf. Chapter 3 of Liu Yangjie's PhD thesis in Nanyang Technological University(NTU) Library.}.

\begin{figure*}[hc]
\includegraphics[width=1.\textwidth]{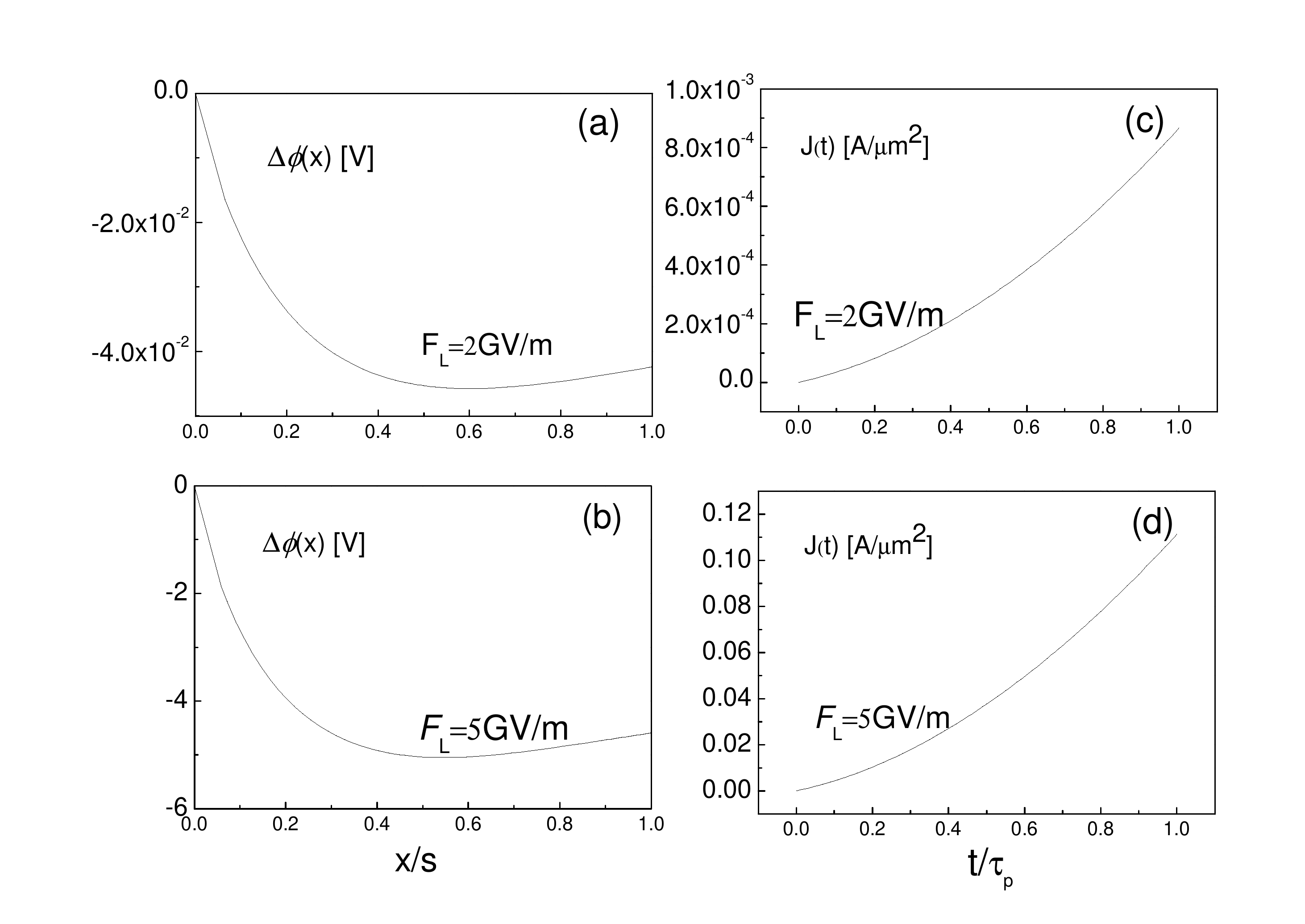}

\caption { \label{fig:Fig2} 
The space charge potential profile [$\Delta\phi (0\leq x/s \leq 1)$] at two different laser field $F_L$ = 2 GV/m (a) and 5 GV/m (b).
The corresponding time-dependent emitting current density $J(t)$ is plotted in panels (c) and (d).
The parameters used are the same as those in figure 1.
}

\end{figure*}

In figure 2, we plot the space charge potential $\Delta\phi(x)$ in the region where electrons distribute $0 < x/s < 1$ for $F_L$ = 2 GV/m [figure 2(a)] 
and  5 GV/m [figure 2(b)] based on the same parameters in figure 1.
It is clear that the space charge effect can be ignored for $F_L \simeq 2 GV/m$, where the magnitude of $\Delta\phi(x)$ is much less than than the one based on $F_L$ = 5 GV/m by 2 orders of magnitude. 
The corresponding emitting current density is plotted in panels 2(c) and 2(d), respectively. This figure confirms our observation above that the critical point for the space-charge effect to dominate is around $\sim$ 5 $\rm GV/m$ under the physical situation investigated in figure 1.

\begin{figure*}[ch]
\includegraphics[width=1.1\textwidth]{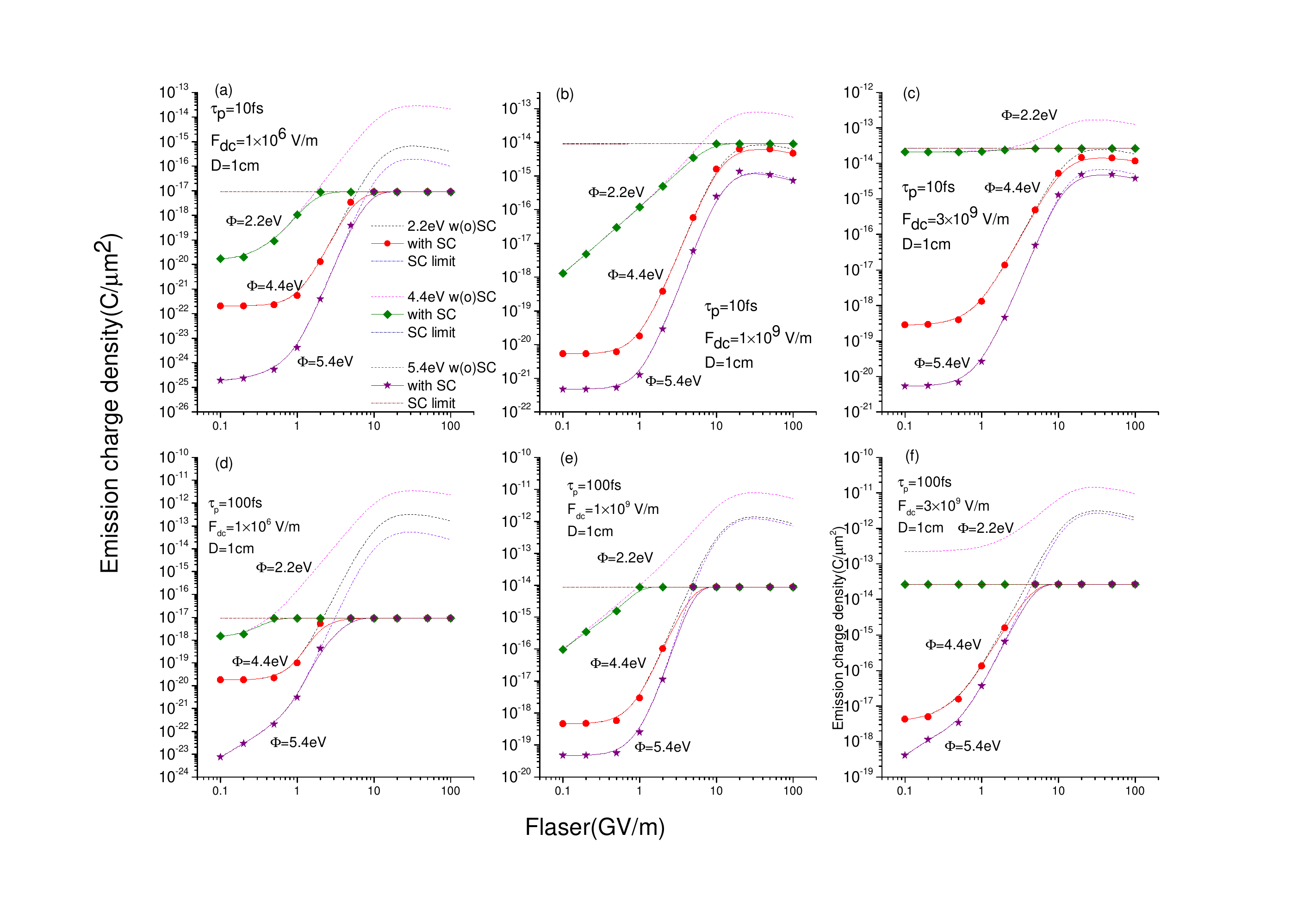}

\caption { \label{fig:Fig3} (Color online) (a-f) 
Emission charge densities $\sigma_{EC}$ as a function of laser field $F_L$ = 0.1 to 100 GV/m at $\tau$ = 10 fs (top) and 100 fs (bottom) for
for different dc fields $F_{dc}$ = 1 MV/m, 1 GV/m and 3 GV/m (left to right).
In each panel, the calculations are performed for $\Phi$ = 2.2 eV (green symbols), 4.4 eV (red symbols) and 5.4 eV (purple symbols). 
(The gap spacing is fixed at $D$  = 1 cm at all cases and the legend in panel (a) applies to (b-f) also. B-spline is taken to smoothen the curves. ) 
}

\end{figure*}

In figure 3, we study the dependence of our results by varying the work function $\Phi$, dc field $F_{\rm dc}$ and laser pulse length $\tau_{\rm p}$. Here, the gap spacing is fixed at $D$ = 1 cm.
In figure 3(a-c), we present the cases at $\tau_{\rm p}$ = 10 fs for $F_{\rm dc}$ = 1 MV/m, 1 GV/m and 3 GV/m (left to right). 
On each panel of figure 3, we show three calculations at different work functions $\Phi$ = 2.2, 4.4 and 5.4 $\rm eV$.
Figure 3(d-f) are presented similarly to figure 3(a-c), except at a longer laser pulse $\tau_{\rm p}$ = 100 $\rm fs$.
By comparing the six panels of figure 3, we make the following observations and discussions.

From the results of figure 3, we can see that higher work function (like 5.4 eV) will enter space charge limited (SCL) regime at a higher laser field. However, the difference between 5.4 eV and 4.4 eV is not significant, both cases of which will be in the SCL regime at $F_L >$ 5 GV/nm. Thus, we expect that experiments using gold with a higher work function (5.4 eV) than tungsten (4.4 eV) will have a less space charge effect at a fixed laser field.

As the laser field increases from $F_L$ = 0.1 to 100 ${\rm GV/m}$, the emission charge density $\sigma_{EC}$ increases within the small $F_L$ range and it gradually saturates to the SCL regime at the high $F_L$ range as expected.
The critical value of $F_L$ that space charge effect becomes important (thus also the threshold to reach space-charge limit) is dependent on all three parameters $F_{dc}$, $\Phi$ and $\tau_p$.
In general, it is easy to reach the SCL regime (critical value of $F_L$ is small) for small $\Phi$, and low $F_{dc}$. 
At small $\Phi$, it is easy to have a large emission current at a fixed field, so space charge limit may also be reached at low $F_L$ values, naturally consistent with our expectation because lower work function makes easier the tunnelling of electrons. 
For example, the critical value to reach the the SCL regime (show in figure 3a) is about $F_L$ 1 to 2 GV/m for $\Phi$ = 2.2 eV as compared to
$F_L$ = 5 to 10 GV/m for $\Phi$ = 4.4 to 5.4 eV.

At small $F_{dc}$ and fixed $D$, SCL current is small, so it is easy to reach the SCL regime for a given emission current.
As $F_{dc}$ increases, emission is greatly promoted because tunnelled probability increases shown in \eref{newV}. 
However, the SCL current density also increases with $F_{\rm dc}$.
For example, the critical value to reach the SCL regime ($\Phi$ =2.2 eV case) is about $F_{\rm L}$ = 1 GV/m (at $F_{dc}$ = 1 MV/m in figure 3(a)) as compared with $F_{\rm L}$ = 2  GV/m (at $F_{dc}$ = 3 GV/m in figure 3(c)).
Thus the dependence on the dc field is not as sensitive as the work function.

The total amount of the SCL charge density $\sigma_{EC}$ (within the laser pulse duration) is found to be nearly identical for various $\tau_p$ = 10 to 100 fs (see the blue dashed lines labelled SC limit).
While the SCL charge current density is expected to increase with small $\tau_p$ according to both short pulse \cite{Valfells2002} and our model here, but the total amount of SCL emitted charge integrated over the small laser pulse length will cancel the short pulse enhancement on current density.
This is understandable as the total amount of charge emitted under SCL will depend on the surface electric field on the cathode \cite{Zhu2013}. 
The critical value for the laser field to reach the space-charge limit will be lower at longer pulse duration of the electron flow. 

Finally, It is important to note that the SCL current density calculated here is based on a 1D model that does not account for the sharpness of the tip as mentioned in the Introduction.
It is expected that the value to be enhanced by a factor of 20 to 100 for a very sharp tip (see figure 8 in \cite{Sun2012}).
Base on this, if we assume the SCL current density is enhanced by a factor of 50, then the threshold of the laser field to reach the SCL current may be increased from 5 GV/m (for a flat surface in figure 1) to about 10 GV/m (for a sharp tip) in order to reach the higher SCL current density.
Note this estimation is only a zero-order approximation, and accurate results will require a protrusive SCL current model beyond the scope of this paper.

\section{Summary}

In summary, we have developed a 1D model to study the space-charge-limited (SCL) emission under the ultrafast laser induced electron emission due to multiphoton absorption and non-equilibrium laser heating on a flat metal surface. 
Our model indicates that the space-charge effect may {\itshape not} be negligible for the high laser field and the threshold to reach the SCL regime is determined for various parameters such as work function, dc applied field and laser pulse length.
Smooth transition from the source-limited emission to SCL emission is obtained.
The calculated results are compared with PIC simulation and a short pulse SCL current model \cite{Valfells2002}.
It is found that the space charge effect is more dominant for materials with a lower work function like tungsten (4.4 eV) as compared with gold (5.4 eV). However for a flat surface, both materials will reach the space charge limited regime at sufficiently high laser field such as $>$ 5 GV/m with a laser pulse length of tens to a hundred femtoseconds.

\ack
The \texttt{VORPAL} simulation work herein was performed in Axle and Fuji servers of A-Star Computational Resource Centre(A*CRC), Singapore, supported by Institute of High Performance Computing, Agency for Science, Technology and Research(IHPC, A*STAR), Singapore.  Some \texttt{C++} code was run under High Performance Computing Centre, Nanyang Technological University(HPC, NTU). Part of this project was benefitted from L Y's 2011 visit to Hong Kong University of Science and Technology.  
This project was funded by the Singapore MOE Tier 2 grant (MOE2008-T2-01-033) and USA AFOAR AOARD grant (11-4069). L Y thanks Wu Lin, Christine Roark, Koh Wee Shing, Chen Shih-Hung and Liu Yao-Li for enlightening guidance for this work, Liang Shijun for his assistance to run the relevant code, and also the considerate editorial work of Editor M Johnston.


\section*{Author contributions}
L K A proposed this starting idea including the iterative numerical algorithm and supervised the whole work. L Y developed the numerical algorithm, performed the calculation and \texttt{VORPAL} simulation, plotted all the graphs from \texttt{Origin}, and wrote the manuscript draft. L K A revised the manuscript.


\newpage
\section*{References}

\bibliographystyle{unsrt}

\begin{thebibliography}{10}



\bibitem{RichEEHB}
Richardson O W 1921
\newblock {\em The Emission of Electricity from Hot Bodies}
\newblock (New York: Longmans, Green)

\bibitem{Fow1928}
Fowler R H and Nordheim L 1928 
Electron emission in intense electric fields
{\em Proc.\ R.\ Soc.\ London, Ser.\ A} {\bf 119} 173

\bibitem{Fow31}
Fowler R H 1931
\newblock The analysis of photoelectric sensitivity curves for clean metals at
  various temperatures
\newblock {\em Phys. Rev.} {\bf 38} 45

\bibitem{Dub32}
DuBridge L A 1932
\newblock A further experimental test of Fowler's theory of photoelectric
  emission
\newblock {\em Phys. Rev.} {\bf 39} 108

\bibitem{Dub33}
DuBridge L A 1933
\newblock Theory of the energy distribution of photoelectrons.
\newblock {\em Phys. Rev.} {\bf 43} 727

\bibitem{Jen07AIEP}
Jensen K L 2007
\newblock {\em Advances in Imaging and Electron Physics}
\newblock (Singapore: Elsevier)

\bibitem{Jen02}
Jensen K L, O'Shea P G and Feldman D W 2002
\newblock Generalized electron emission model for field, thermal and
  photoemission
\newblock {\em Appl. Phys. Lett.} {\bf 81} 3867

\bibitem{Child1911}
Child C D 1911
\newblock Discharge from hot Cao 
\newblock {\em Phys. Rev.}(Series I) {\bf 32}(5) 492--511

\bibitem{Langmuir1913}
Langmuir I 1913
\newblock The effect of space charge and residual gases on thermionic currents
  in high vacuum
\newblock {\em Phys. Rev.} {\bf 2}(6) 450--486

\bibitem{Lau94}
Lau Y Y, Liu Y and Parker R K 1994
\newblock Electron emission: From the Fowler-Nordheim relation to the
  Child-Langmuir law
\newblock {\em Phys. Plasmas} {\bf 1} 2082

\bibitem{Koh06}
Koh W S and Ang L K 2006
\newblock Transition of field emission to space-charge-limited emission in a
  nano gap
\newblock {\em Appl. Phys. Lett.} {\bf 89} 183107

\bibitem{Valfells2002}
Valfells {\'A}, Feldman D W, Virgo M, O'Shea P G and Lau Y Y 2002
\newblock Effects of pulse-length and emitter area on virtual cathode formation
  in electron guns
\newblock {\em Phys. Plasmas} {\bf 9}(5) 2377--2382

\bibitem{Ang2007}
Ang L K and Zhang P 2007
\newblock Ultrashort-pulse Child-Langmuir law in the quantum and relativistic
  regimes
\newblock {\em Phys. Rev. Lett.} {\bf 98}(16) 164802

\bibitem{Hommelhoff2006a}
Hommelhoff P, Sortais Y, Aghajani-Talesh A and Kasevich M A 2006
\newblock Field emission tip as a nanometer source of free electron femtosecond
  pulses
\newblock {\em Phys. Rev. Lett.} {\bf 96}(7) 4

\bibitem{Hommelhoff2006b}
Hommelhoff P, Kealhofer C and Kasevich M A 2006
\newblock Ultrafast electron pulses from a tungsten tip triggered by low-power
  femtosecond laser pulses.
\newblock {\em Phys. Rev. Lett.} {\bf 97}(24) 4

\bibitem{Ropers2007}
Ropers C, Solli D R, Schulz C P, Lienau C and Elsaesser T 2007
\newblock Localized multiphoton emission of femtosecond electron pulses from
  metal nanotips 
\newblock {\em Phys. Rev. Lett.} {\bf 98}(4) 043907

\bibitem{Barwick2007}
Barwick B, Corder C, Strohaber J, Chandler-Smith N, Uiterwaal C and
  Batelaan H 2007
\newblock Laser-induced ultrafast electron emission from a field emission tip
\newblock {\em New J. Phys.} {\bf 9} 142

\bibitem{Wu2008}
Wu L and Ang L K 2008
\newblock Nonequilibrium model of ultrafast laser-induced electron photofield
  emission from a dc-biased metallic surface
\newblock {\em Phys. Rev.} B {\bf 78}(22) 224112

\bibitem{Sch2010}
Schenk M, Kr\"{u}ger M and Hommelhoff P 2010
\newblock Strong-field above-threshold photoemission from sharp metal tips.
\newblock {\em Phys. Rev. Lett.} {\bf 105} 257601

\bibitem{Yanagisawa2011}
Yanagisawa H, Hengsberger M, Leuenberger D, Kl\"ockner M, Hafner C, Greber T and Osterwalder J 2011
\newblock Energy distribution curves of ultrafast laser-induced field emission
  and their implications for electron dynamics
\newblock {\em Phys. Rev. Lett.} {\bf 107}(8) 087601

\bibitem{Yalunin2011}
Yalunin S,  M and Ropers C 2011
\newblock Strong-field photoemission from surfaces: Theoretical approaches.
\newblock {\em Phys. Rev.} B {\bf 84}(19) 195426

\bibitem{Pant2012}
Pant M and Ang L K 2012
\newblock Ultrafast laser-induced electron emission from multiphoton to optical
  tunneling
\newblock {\em Phys. Rev.} B, {\bf 86}(4) 045423

\bibitem{Ang2013}
Ang L K and Pant M 2013
\newblock Generalized model for ultrafast laser induced electron emission from
  a metal tip
\newblock {\em Phys. Plasmas} {\bf 20}(5) 056705--6

\bibitem{Bormann2010}
Bormann R, Gulde M, Weismann A, Yalunin S and Ropers C 2010
\newblock Tip-enhanced strong-field photoemission
\newblock {\em Phys. Rev. Lett.} {\bf 105} 147601

\bibitem{Kruger2011}
Kr\"{u}ger M, Schenk M and Hommelhoff P 2011
\newblock Attosecond control of electrons emitted from a nanoscale metal tip
\newblock {\em Nature} {\bf 475}(7354) 78--81

\bibitem{Herink2012}
Herink G, Solli D R, Gulde M and Ropers C 2012
\newblock Field-driven photoemission from nanostructures quenches the quiver
  motion.
\newblock {\em Nature} {\bf 483} 190-193

\bibitem{Park2012}
Park D J, Piglosiewicz B, Schmidt S, Kollmann H, Mascheck M and Lienau C 2012
\newblock Strong field acceleration and steering of ultrafast electron pulses
  from a sharp metallic nanotip 
\newblock {\em Phys. Rev. Lett.} {\bf 109}(24) 244803

\bibitem{Mueller2013}
Mueller B Y and Rethfeld B 2013
\newblock Relaxation dynamics in laser-excited metals under nonequilibrium
  conditions.
\newblock {\em Phys. Rev.} B {\bf 87}(3) 035139

\bibitem{Pant2013}
Pant M and Ang L K 2013
\newblock Time-dependent quantum tunneling and nonequilibrium heating model for
  the generalized Einstein photoelectric effect
\newblock {\em Phys. Rev.} B {\bf 88}(19) 195434

\bibitem{Wendelen2010}
Wendelen W, Autrique D and Bogaerts A 2010
\newblock Space charge limited electron emission from a Cu surface under
  ultrashort pulsed laser irradiation
 \newblock   {\em Appl. Phys. Lett.} {\bf 96}(5) 051121 


\bibitem{Wendelen2012}
Wendelen W, Mueller B Y, Autrique D, Rethfeld B and Bogaerts A 2012
\newblock Space charge corrected electron emission from an aluminum surface
  under non-equilibrium conditions
\newblock {\em J. Appl. Phys.} {\bf 111}(11) 113110

\bibitem{Lau2001}
Lau Y Y 2001
\newblock Simple theory for the two-dimensional Child-Langmuir law
\newblock {\em Phys. Rev. Lett.} {\bf 87}(27) 278301

\bibitem{Nieter2004}
Nieter C and Cary J R 2004
\newblock {VORPAL}: a versatile plasma simulation code
\newblock {\em J. Comput. Phys.} {\bf 196}(2) 448--473

\bibitem{Mahalingam2009}
Mahalingam S, Nieter C, Loverich J, Smithe D and Stoltz P 2009
\newblock Space charge limited currents calculations in coaxial cylindrical
  diodes using particle-in-cell simulations 
\newblock {\em The Open Plasma Physics Journal} {\bf 2} 7

\bibitem{Chen2011}
Chen S H, Tai L C, Liu Y L, Ang L K and Koh W S 2011
\newblock Two-dimensional electromagnetic Child--Langmuir law of a short-pulse
  electron flow
\newblock {\em Phys. Plasmas} {\bf 18}(2) {023105}

\bibitem{Zhu2013}
Zhu Y B, Zhang P, Valfells {\'A}, Ang L K and Lau Y Y 2013
\newblock Novel scaling laws for the Langmuir-Blodgett solutions in cylindrical
  and spherical diodes
\newblock {\em Phys. Rev. Lett.} {\bf 110}(26) 265007

\bibitem{Sun2012}
Sun S and Ang L K 2012
\newblock Onset of space charge limited current for field emission from a
  single sharp tip
\newblock {\em Phys. Plasmas} {\bf 19}(3) 033107

\bibitem{Liu2011}
Liu Y, Wu L and Ang L K June-17 2011
\newblock Saturation of Ultrafast Laser Excited Electron Emission from Metallic Nanotip at High Current Regime
\newblock presented at {\em East Asian Postgraduate Workshop on Nanoscience and Technology}, The Hong Kong University of Science and Technology (HKUST), {\bf 109-110}; Liu Y and Ang L K May-22 2013
\newblock Space Charge Effect of Time-dependent Emission Current Excited from Ultrafast Laser
\newblock presented at {\em 14th IEEE International Vacuum Electronics Conference (IVEC)}, Paris, Poster Session {\bf I} 


	

\end{thebibliography}

\newpage


\end{document}